\documentclass{emulateapj}

\shorttitle{}
\shortauthors{H.\ Wu, et al.}

\received{}
\accepted{}

\begin{document}

\title{
PAH AND MID-IR LUMINOSITIES AS MEASURES OF STAR-FORMATION RATE IN
{\it SPITZER FIRST LOOK SURVEY} GALAXIES
}

\author{
HONG WU\altaffilmark{1,2},
CHEN CAO\altaffilmark{1},
CAI-NA HAO\altaffilmark{1},
FENG-SHAN LIU\altaffilmark{1},
JIAN-LING WANG\altaffilmark{1,3},
XIAO-YANG XIA\altaffilmark{3},
ZU-GAN DENG\altaffilmark{4},
C.\ KE-SHIH YOUNG\altaffilmark{1,5}
}

\email{hwu@bao.ac.cn}
\altaffiltext{1}{National Astronomical Observatories, Chinese Academy of Sciences, Beijing 100012, P.R.\ China}
\altaffiltext{2}{Visiting Scholar, Harvard-Smithonian Center for Astrophysics, 60 Garden Street,  Cambridge, MA 02138}
\altaffiltext{3}{Department of Physics, Tianjin Normal University, Tianjin 300074, P.R.\ China}
\altaffiltext{4}{College of Physical Sciences, Graduate School of the Chinese Academy of Sciences, P.O.\ Box 3908, Beijing 100039, P.R.\ China}
\altaffiltext{5}{Department of Physics, University of Hong Kong, Pokfulam Road, Hong Kong, P.R.\ China}

\begin{abstract}
We present evidence that the 8~$\mu$m (dust) and 24~$\mu$m luminosities of star-forming
galaxies are both strongly correlated with their 1.4~GHz and H$\alpha$ luminosities over a
range in luminosity of two-to-three orders of magnitude. At the bright end, the correlations
are found to be essentially linear over a luminosity range of about two orders of magnitude
(corresponding to star-formation rates of several-tenths to several tens of solar masses per
year). However, at the faint end, there appears to be a slope change for dwarf
galaxies, possibly due to the lower dust-to-gas ratios and lower metallicities of the dwarfs.
The correlations suggest that PAH features and mid-IR continuum emissions are good measures
of the star formation rates of galaxies, and we present calibrations of star-formation rates
based on existing radio and H$\alpha$ relations. Our findings are based on a sample of
star-forming galaxies selected from the main field of the Spitzer First Look Survey with the
aid of spectroscopic data from the Sloan Digital Sky Survey and VLA 1.4~GHz data. 
\end{abstract}


\section{INTRODUCTION}
\label{sec intro}

Star formation rates (SFRs) are fundamental measures of galaxy formation and evolution. 
Reliable measurement of SFRs could provide us with important clues and constraints 
concerning the evolution of galaxies. With the less extinction and the less 
contamination from cirrus, the mid-infrared (MIR) might be ideal tracer of star 
formation.  In addition, broad emission MIR features, such as 
3.3, 6.2, 7.7, 8.6, 11.2, 12.7, 16.4, 17.1~$\mu$m, have already been found to be 
present in most star-forming galaxies by both the Infrared Space Observatory (ISO, 
Kessler et al.\ 1996; Sturm et al.\ 2000) and Spitzer Space Telescope (Smith et al.\ 2004), 
and these features are thought to be related to the vibrational emissions of polycyclic 
aromatic hydrocarbons (PAHs).

The ISO 6.75~$\mu$m band just covers the PAH features whilst the 15~$\mu$m band just
covers the dust continuum from very small grains (VSGs). However, it turns out that PAH
emission from the 12.7~$\mu$m feature can also appear in the 15~$\mu$m band, so observed 
15~$\mu$m fluxes will not be due to VSG continua exclusively. Elbaz et al.\ (2002) have 
demonstrated a relationship between MIR (both 6.75$~\mu$m and 15$~\mu$m) and 
IR luminosities; whilst Roussel et el.\ (2001) and F\"{o}rster-Schreiber et al.\
(2004) have demonstrated one between MIR (both 6.75$~\mu$m and 15$~\mu$m) and
H$\alpha$ luminosities in spiral disks. 

With the launch of the Spitzer Space Telescope (Werner et al.\ 2004), it become available 
to determine SFRs of millions of galaxies in the MIR bands. The IRAC~8$\mu$m band covers the 
strongest 7.7~$\mu$m PAH feature, whilst the MIPS 24$\mu$m band avoids most of the PAH 
features and can sample the continuum emission of VSGs better than the ISO~15$\mu$m band can.  
Calzetti et al.\ (2005) first obtained the tight relations between Spitzer MIR (8$~\mu$m and 
24$~\mu$m) and P$\alpha$ luminosities based on the new Spitzer observations of the 42 knots 
on M51. The goal of our work is to examine whether Spitzer Space Telescope IRAC 8~$\mu$m 
and MIPS~24$\mu$m luminosities can be used to measure SFRs of whole galaxies through 
comparisons with known SFR indicators such as H$\alpha$ and radio luminosities. However, 
since the MIR continuum can easily be contaminated by emissions from dust heated by AGNs, 
special care is needed to avoid galaxies with AGNs.

In Section 2, we describe how we define our sample of star-forming galaxies.
Correlation analysis between MIR luminosities and radio/H$\alpha$ luminosities, 
and calibrations of MIR luminosities to SFRs are presented in Section 3 and 4. 
Throughout this paper, we adopt a cosmology with $\Omega_M=0.3$, $\Omega_{\Lambda}=0.7$, and $H_0=70$.

\section{SAMPLE SELECTION}
Our galaxy sample was taken from the extragalactic component of the main field of the Spitzer
First Look Survey (FLS).
In particular, it was drawn from the overlap area of about 3.7 degree$^2$ of the main field
that has been imaged by both instruments: IRAC (3.6, 4.5, 5.8, 8.0~$\mu$m) and MIPS
(24$\mu$m). This area has also been covered by the SDSS (Stoughton et al.\ 2002). The IRAC
images (in all four IRAC bands) were mosaiced from the ``Basic Calibrated Data'' (BCD) 
as described by Fazio, et al.\ (2004) and Huang et al.\ (2004); whilst the MIPS images were 
mosaiced from post-BCD images. Matching the sources detected 
by SExtractor (Bertin \& Arnouts 1996) in all five bands with the 2mass sources 
ensured astrometric uncertainties of less than 0.1$\arcsec$. The IRAC and MIPS bands have flux 
calibration accuracies of better than 10\% (Fazio et al.\ 2004; Rieke et al.\ 2004).

FLS sources were then matched with the main galaxy sample of the second data release of the
SDSS (Strauss et al.\ 2002) with a matching radius of 2$\arcsec$.  The galaxies with  
H$\alpha$ fluxes of S/N~$\ge$~5 were selected.  Based on the traditional line-diagnostic
diagram [NII]/H$\alpha$ versus [OIII]/H$\beta$ (Veilleux and Osterbrock 1987; Wu et al.\ 1998),
narrow-line AGNs were removed. The fluxes of all of the emission lines were taken from the SDSS
catalogue of measured line fluxes (Tremonti et al.\ 2004, version 5.0$\_$4). We also included 
1.4~GHz fluxes when the coordinates of VLA 1.4~GHz radio catalogue of Condon et al.\ (2003)
matched the sample galaxies to an accuracy of 0.5$~\arcsec$.

Our sample of the star-forming galaxies contains a total of 91 galaxies with both H$\alpha$ and MIR
fluxes in all four IRAC bands. 70 of these also have MIPS 24$\mu$m fluxes. Due to sensitivity
limitations affecting the VLA observations (Condon et al.\ 2003), only 38 of our 91 IRAC sources and
33 of our 70 MIPS sources have 1.4~GHz fluxes.  All these objects are local with redshifts of less than 0.2. 
There are 12 dwarf galaxies with absolute $B$ magnitudes fainter than $-$18.0 in our sample, 
but only three of these dwarfs have radio data.

\section{MIR LUMINOSITIES AND CORRELATION ANALYSIS}
\label{radio_ha}

To examine whether MIR 8$\mu$m and 24$\mu$m luminosities can be taken as measures of the
SFRs of star-forming galaxies, we first compare them with 1.4~GHz and H$\alpha$ luminosities,
which have already been extensively used as measures of SFRs.
To correct for redshift effects, we adopt a steep non-thermal radio spectrum whose flux obeys a
power law of the form $\nu^{-\alpha}$ (Cardiel et al.\ 2003) and has a spectral index of $\alpha$ = 0.8
(Condon 1992), and compute rest-frame radio luminosities at 1.4 GHz.

As for the H$\alpha$ luminosities,  both extinction and aperture corrections were applied to
derive the total H$\alpha$ emission of the whole galaxy based on the fibre spectra.
The extinction correction included both the foreground extinction of the Milky Way and the
intrinsic extinction of galaxy itself. The Milky Way extinction was corrected by the 
parameterized curve of Cardelli et al.\ (1989) and the intrinsic extinction $A_{H\alpha}$ 
was obtained from the Balmer decrement $F_{H\alpha}/F_{H\beta}$ (Calzetti 2001). For the 
aperture corrections, we followed the Formula~A1 of Hopkins et al.'s (2003). The corrections
were based on the $r$ band fiber magnitude and the corresponding Petrosian magnitude, which can 
be approximately treated as the total flux of the whole galaxies in $r$ band.

Using SExtractor, ``AUTO" magnitudes were obtained for the total luminosities of the sample galaxies
in the IRAC and MIPS 24$\mu$m bands. We adopted the SED of the normal HII galaxy NGC~3351 from IRS observation
of the Spitzer Legacy Program SINGS (Kennicutt et al.\ 2003) as the template for the K-corrections for 
our sample galaxies in both the 24$\mu$m and 8$\mu$m bands. To estimate the uncertainty of SED template, 
we compared it with those of other 20 SINGS normal galaxies, M82, and the models of normal and starburst galaxies of
Lagache, Dole and Puget (2003). Within redshift of 0.2, the difference of K-correction is at most 30\%.
Although PAH emissions dominate the 8.0$\mu$m band for most of the sample galaxies, there is still
a stellar continuum in this band, especially for the galaxies of earlier type. To remove the stellar
contribution in this band, we made use of the 3.6$\mu$m band which avoids the strong PAH emission
and used it to estimate the stellar contribution in the 8.0$\mu$m band. We selected 74 galaxies 
without optical emission lines in the FLS and assume they are dust-free at 3.6$\mu$m.
The statistics show that the averaged magnitude differences between the 3.6$\mu$m and
8.0$\mu$m bands for these galaxies is about 1.45~mag with a scatter of 0.28~mag for $z < 0.2$. 
This corresponds the factor of 0.26 to scale the stellar continuum of 3.6$\mu$m to that of 8$\mu$m,
and a little larger than the 0.232 of Helou et al.\ (2004). This factor was then used to 
subtract the stellar continuum contributions from the estimated luminosities for sample galaxies 
in the 8$\mu$m band. Generally, the stellar continuum
flux amounted to about 10\% of the 8.0$\mu$m flux, similar to that of Engelbracht et al.\ (2005). 
The resulting luminosities which represent the dust emissions of our sample galaxies will henceforth 
be denoted 8$\mu$m(dust). Though the PAH emissions dominate the 8$\mu$m band, but the dust grains with 
size $>$ 50$\rm \r{A}$ can still contribute to the 8$\mu$m continuum at some level (Li \& Draine 2001).

\begin{figure}
\centerline{\includegraphics[height=0.9\textwidth]{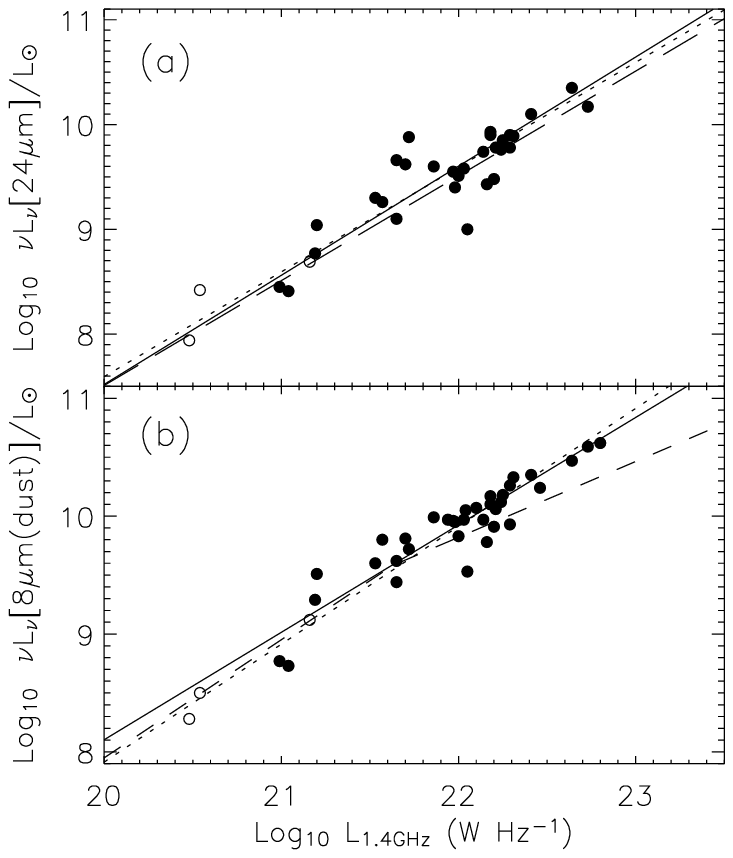}}
\caption{The correlations between the MIR and radio luminosities.  The filled circles are 
the sample galaxies and  the unfilled symbols are the dwarf galaxies with $M_B$ $< -18.0$~mag. 
The solid and dotted lines are the best non-linear and linear fits respectively for 
star-forming galaxies. (a) 24$\mu$m versus radio (1.4~GHz). The long-dashed line represents 
the linear fit to the radio and 24$\mu$m luminosities obtained by Appleton et al.\ (2004) 
for radio-quiet sources with K-corrections based on an M82 template. (b) 8$\mu$m(dust) 
versus radio. The short-dashed line represents the relation of the radio and 6.75$\mu$m (ISO)
luminosities drawn from Elbaz et al.\ (2002). 
} \label{fig1}
\end{figure}

Figure~\ref{fig1} shows (a) 24$\mu$m and (b) 8$\mu$m(dust) luminosity versus the 1.4 GHz
radio luminosity in the log-log space. We can see that a good correlation exists between the MIR and
radio luminosities over a luminosity range of about 2 orders of magnitudes. The unfilled symbols
represent the only three dwarf galaxies with catalogued radio fluxes. These three dwarfs seem to
follow similar correlations to those found for brighter galaxies. The solid lines in the figures are the
best fits to all brighter galaxies with two-variable regression. The dotted lines are the linear fits.
The fitted parameters and the correlation coefficients are listed in Table~1. 
As evident from Figure~\ref{fig1}(a), our results are consistent with those of Appleton et al.\ (2004). 
However, our correlation exhibits a greatly reduced scatter of 0.17 cf.\ Appleton et al.'s (2004) 0.27. 
In the case of Figure~\ref{fig1} (b), our results appear to be consistent
with those of Elbaz et al.\ (2002) at the faint end but there is a departure at the bright end.
The correlation was deduced from the Elbaz et al.'s relation of the bolometric IR luminosity versus
ISO 6.75$\mu$m luminosity and the radio "$q$" parameter. 

\begin{figure}
\centerline{\includegraphics[height=0.9\textwidth]{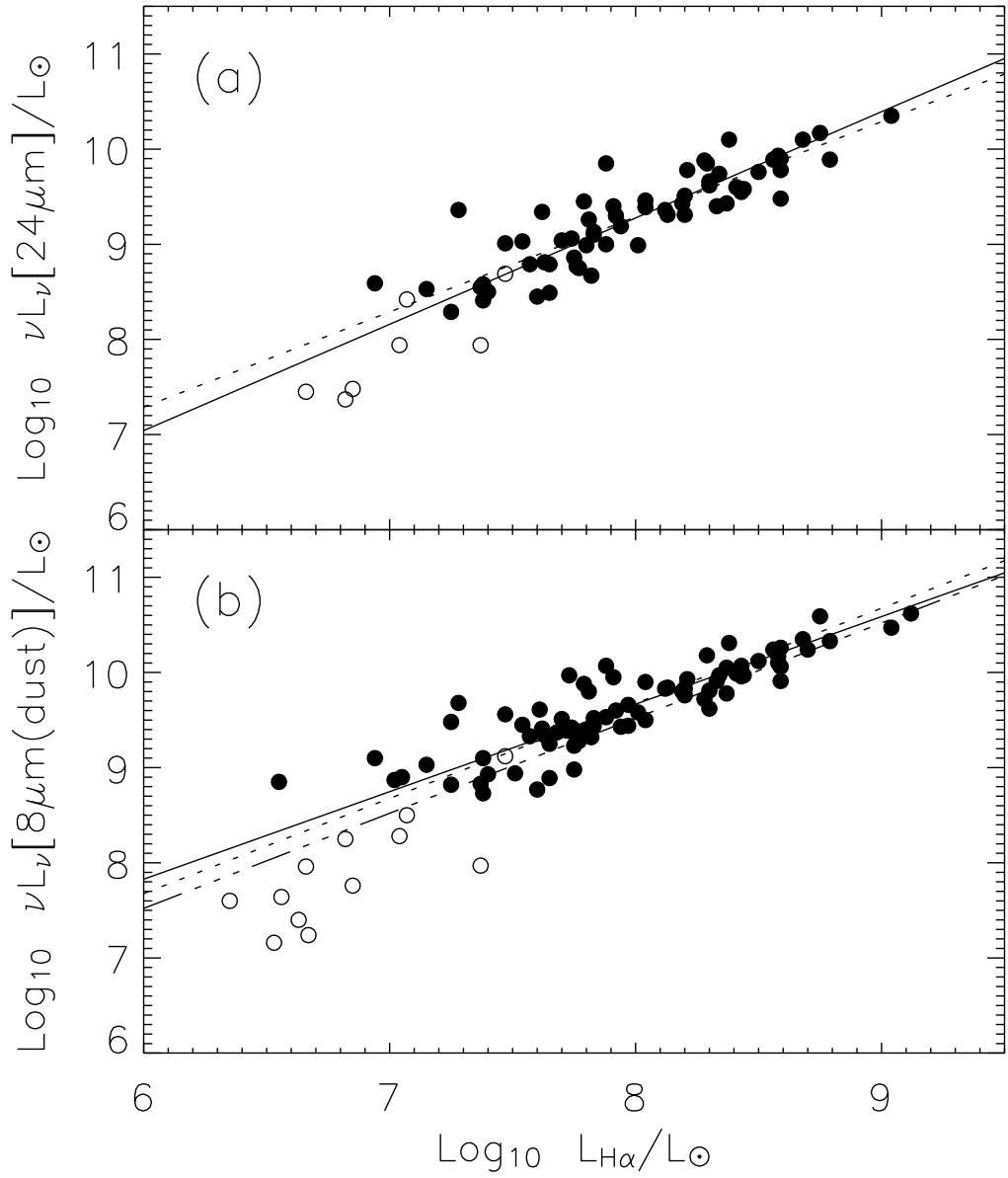}}
\caption{The correlations between the MIR and H$\alpha$ luminosities.
The filled, unfilled circles, the solid and dotted lines are as in Figure~1.
(a) 24$\mu$m versus H$\alpha$.  (b) 8$\mu$m(dust) versus H$\alpha$.  The dot-dashed line is
the linear relation between ISO 6.75$\mu$m and H$\alpha$ luminosities for spiral disks found
by Roussel et al.s (2001).
} \label{fig2}
\end{figure}

Figure~\ref{fig2} shows (a) 24$\mu$m and (b) 8$\mu$m(dust) luminosity versus H$\alpha$ luminosity in
log-log space. We can observe a good correlation between these two MIR luminosities and the H$\alpha$
luminosity for the sample galaxies except that there appears to be a slope change for dwarf galaxies.
Both of the linear correlations for non-dwarf galaxies have correlation coefficients of about 0.9.
As evident from Figure~\ref{fig2} (b) broad agreement is found with the results of Roussel et al.\ (2001)
which were based on a straight-line fit to 6.75$\mu$m (ISO) and H$\alpha$ luminosities for spiral disks.

\section{CALIBRATION OF 24$\mu$m AND 8$\mu$m(dust) STAR FORMATION RATES}
\label{subsec pah}

From the strengths of the correlations listed in Table~1, there appears to be a strong link between both
MIR 24$\mu$m and 8$\mu$m(dust)  luminosity and both 1.4~GHz radio and H$\alpha$ luminosity.
As both radio and H$\alpha$ luminosities are already considered to be good
measures of SFR, we can estimate SFRs based on MIR (24$\mu$m and 8$\mu$m(dust)) luminosities
from either radio and/or H$\alpha$ SFRs.  

Combined with the linear relations in Table~1 and the SFR-radio luminosity relation given by Yun, Reddy and 
Condon (2001), we can derive SFRs from both the 24$\mu$m and 8$\mu$m(dust) luminosities:
\begin{equation}
\label{sfr24_vla}
SFR_{24\mu\,m}(M_{\odot} yr^{-1}) = \frac{\nu\,L_{\nu}[24\mu\,m]}{6.66\times 10^{8} L_{\odot}},
\end{equation}
\begin{equation}
SFR_{8\mu\,m(dust)}(M_{\odot} yr^{-1}) = \frac{\nu\,L_{\nu}[8\mu\,m(dust)]}{1.39\times 10^{9} L_{\odot}}.
\end{equation}

Similarly, we can also derive SFRs from H$\alpha$ luminosities based on Kennicutt's (1998) SFR formula:
\begin{equation}
SFR_{24\mu\,m}(M_{\odot} yr^{-1}) = \frac{\nu\,L_{\nu}[24\mu\,m]}{6.43\times 10^{8} L_{\odot}},
\end{equation}
\begin{equation}
\label{sfr8_ha}
SFR_{8\mu\,m(dust)}(M_{\odot} yr^{-1}) = \frac{\nu\,L_{\nu}[8\mu\,m(dust)]}{1.57\times 10^{9} L_{\odot}}.
\end{equation}
However, due to the change in slope exhibited by the dwarf galaxies in Figure~\ref{fig2}, the latter two
relations are necessarily limited to the normal star-forming galaxies.

There are many factors which could affect calibrations and increase the scatter in the
correlations. These include the accuracy of the fibre-aperture corrections, the validity of using the Balmer
decrement to estimate obscuration in galaxies (Hopkins et al.\ 2003), possible radio and MIR contamination
from obscured weak non-optical AGNs, the template used for K-correction, and so on. 
We have therefore compared the SFRs derived from
formulae (1) through (4). Reassuringly, we found that the agreement between the SFRs based on radio and
H$\alpha$ calibrations are around 10\% for both the 24$\mu$m and 8$\mu$m(dust) relations respectively.
This level of error is well within the 1-$\sigma$ scatters as listed in table 1. 

\section{DISCUSSION}
\label{sec discuss}

Based on the relations we have derived, we suggest that the 24$\mu$m continuum could be due to the
warm dust component associated with star formation regions (O'Halloran et al.\ 2005). This support
the case made by Dale et al.\ (2001) that the 20-42 $\mu$m continuum may be the best dust emission
tracer of current star formation in galaxies. The results of the HII knots in M51 (Calzetti et al.\ 2005)
and the 75 nearby galaxies from SINGS (Dale et al.\ 2005) also suggest that the 24$\mu$m emission is
a useful tracer of the local star formation.  Furthermore, the  agreement between our radio-24$\mu$m
relation for star-forming galaxies and Appleton et al.'s (2004) relation for radio-quiet sources in the same
Spitzer FLS region (Figure~\ref{fig1} (a)), indicates that even for many optical AGNs, the 24$\mu$m emission is
still dominated by star formation. The much larger scatter in the radio-24$\mu$m relation for radio-quiet 
sources could then be explained by the extra contribution from the central AGN.

Though we have drawn the linear calibration from MIR luminosity to SFRs, the best fit
slopes for the 8$\mu$m(dust) luminosity versus H$\alpha$/radio luminosity is obviously shallower than 
those for 24$\mu$m (Table~1). Calzetti et al.\ (2005) obtained a similar results for HII knots in M51.
Their 24$\mu$m and P$\alpha$ luminosities are nearly linear correlated, but the slope for 8$\mu$m 
luminosity versus P$\alpha$ luminosity is much shallower than one.
Peeters, Spoon and Tielens (2004) have pointed out that PAHs may be better tracers of B stars than
tracers of massive star formation (O stars). PAH features can be strong in moderate UV radiation fields
such as photo-dissociation regions, but the PAH carriers could be destroyed in hard radiation field
(Galliano et al.\ 2005). If it is true, 8$\mu$m(dust) flux might not be a linear measure of intensive 
starburst activity, this could explain the shallower slope of 8$\mu$m(dust).  Alternatively, the VSG continuum
might be a useful tracer of star formation in higher UV-radiation fields (F\"{o}rster-Schreiber et al.\ 2004),
though VSGs might also be destroyed in very harsh radiation fields (Contursi et al.\ 2000). 
Since the sample used here are star-forming galaxies, a sample of luminous or even ultraluminous infrared 
galaxies is needed to check if such relations could be extend to intensive starburst environments.

Kewley et al.\ (2002) showed that SFRs derived from H$\alpha$ emission are similar to those derived
from IR emission even at SFR of 0.01 $M_{\odot}$ yr$^{-1}$. However Figure~\ref{fig2} shows that the
dwarf galaxies systematically deviate from the  correlation of the star-forming galaxies to the side of
weak PAH emission in the log-log space. 
Madden (2000) and Dale et al.\ (2005) have also shown that PAH features are weak in dwarf galaxies. 
It is well known that many dwarf galaxies are gas-rich and have higher gas-to-dust ratios 
(as well as lower metallicities) than normal galaxies. 
In low metallicity environments, the UV radiation 
field is presumably strong enough to destroy the PAH carriers and weaken the PAH features, for example, 
Blue compact dwarf galaxies SBS~0335-052 (Houck et al.\ 2004). Estimating the oxygen-abundance with 
the emission line ratios (Kobulnicky, Kennicutt, \& Pizagno 1999), 
we find that two of four dwarf galaxies (including the most deviant one) exhibit much lower metallicities
(about one half or one-third of the solar metallicity) than those of the normal star-forming galaxies.
So the slope of the MIR-SFR relations might therefore be metallicity dependent. This is also supported by 
Engelbracht et al.\ (2005)'s result that the 8$\mu$m-to-24$\mu$m flux ratio strongly depends on metallicity
for a sample of 34 star-forming galaxies.

\section{ACKNOWLEDGEMENTS}
\label{sec acknowledge}

The authors would like to thank Jia-sheng Huang and Zhong Wang for useful discussion and advice.
We gratefully acknowledge the anonymous referee for insightful comments to improve the paper.
This project was supported by the NSF of China through grants No.~10273012, 10333060, 10473013 and
NKBRSF~G1999075404; We thanks the Spitzer Science Center operating the Spitzer Archive.
We also thank the SDSS team for creation and distribution of the SDSS Archive.

\clearpage

\clearpage

\clearpage

\begin{table}
\begin{center}
\caption{Correlation coefficients for luminosity relations for star-forming galaxies}
\label{tab correlation}
\small
\begin{tabular}{llrrrrrr}
\tableline\tableline
$y$   & $x$  & a & b  & r & c  &  N & \\
\tableline
$\nu L_{\nu}[24\mu m](L_{\odot})$ &$ L_{1.4GHz}(W\ Hz^{-1})$ &-13.31$\pm$0.42&1.04$\pm$0.09& 0.82&-12.41$\pm$0.17& 30 \\
$\nu L_{\nu}[8\mu m](L_{\odot})$ &$ L_{1.4GHz}(W\ Hz^{-1})$  &-10.12$\pm$0.31&0.91$\pm$0.07& 0.88&-12.09$\pm$0.13& 35 \\
$\nu L_{\nu}[24\mu m](L_{\odot})$ &$ L_{H\alpha}(L_{\odot})$&  0.34$\pm$0.19&1.12$\pm$0.07& 0.88&  1.29$\pm$0.19& 63 \\
$\nu L_{\nu}[8\mu m](L_{\odot})$ &$ L_{H\alpha}(L_{\odot})$ &  2.31$\pm$0.15&0.92$\pm$0.05& 0.87&  1.68$\pm$0.19& 79\\
\tableline
\end{tabular}\\
\end{center}
{a and b are the coefficients of the non-linear fitting of $\log_{10}(y)=a+b\log_{10}(x)$ and c is the
coefficient of the linear fitting of $\log_{10}(y)=c+\log_{10}(x)$.  $r$ is the Spearman's correlation
coefficient, rho, and $N$ is the number of sample galaxies (excluding the dwarf galaxies) used for the
fitting.}
\end{table}

\label{lastpage}
\end{document}